\documentclass{PoS}
\usepackage{amsmath}

\usepackage{lineno}

\title{The Optical System for the Large Size Telescope of the Cherenkov Telescope Array}

\ShortTitle{The Optical System for the Large Size Telescope of the Cherenkov Telescope Array }

\author{\speaker{M.~Hayashida}$^{a}$, K.~Noda$^{b}$, M.~Teshima$^{a, b}$, U.~Barres de Almeida$^{c}$, M.~Chikawa$^{d}$, N.~Cho$^{e}$, S.~Fukami$^{a}$, A.~Gadola$^{f}$, Y.~Hanabata$^{a}$, D.~Horns$^{g}$, C.~Jablonski$^{b}$, H.~Katagiri$^{e}$, M.~Kagaya$^{e}$, M.~Ogino$^{a}$, A.~Okumura$^{h,i}$, T.~Saito$^{j}$, R.~Stadler$^{b}$, S.~Steiner$^{f}$, U.~Straumann$^{f}$, A.~Vollhardt$^{f}$, H.~Wetteskind$^{b}$, T. ~Yamamoto$^{k}$, T.~Yoshida$^{e}$, for the CTA Consortium\footnote{Full consortium author list at http://cta-observatory.org}\\
        E-mail: \email{mahaya@icrr.u-tokyo.ac.jp}

{\footnotesize
$^{a}$ Institute for Cosmic-Ray Research, the University of Tokyo, Japan;
$^{b}$ Max Planck Institute for Physics, Germany;
$^{c}$ CBPF/MCT, Brasil;
$^{d}$ Kinki University, Japan;
$^{e}$ Ibaraki University, Japan;
$^{f}$ University of Zurich, Switzerland;
$^{g}$ University of Hamburg, Germany;
$^{h}$ Nagoya University, Japan;
$^{i}$ MPI for Nuclear Physics, Germany;
$^{j}$ Kyoto University, Japan;
$^{k}$ Konan University, Japan;}

}

\abstract{The Large Size Telescope (LST) of the Cherenkov Telescope Array (CTA) is designed to achieve a threshold energy of 20\,GeV. The LST optics is composed of one parabolic primary mirror 23\,m in diameter and 28\,m focal length. The reflector dish is segmented in 198 hexagonal, 1.51\,m flat to flat mirrors. The total effective reflective area, taking into account the shadow of the mechanical structure, is about 368\,m$^2$. The mirrors have a sandwich structure consisting of a glass sheet of 2.7\,mm thickness, aluminum honeycomb of 60\,mm thickness, and another glass sheet on the rear, and have a total weight about 47\,kg. The mirror surface is produced using a sputtering deposition technique to apply a 5-layer coating, and the mirrors reach a reflectivity of $\sim$94\% at peak. The mirror facets are actively aligned during operations by an active mirror control system, using actuators, CMOS cameras and a reference laser. Each mirror facet carries a CMOS camera, which measures the position of the light spot of the optical axis reference laser on the target of the telescope camera. The two actuators and the universal joint of each mirror facet are respectively fixed to three neighboring joints of the dish space frame, via specially designed interface plate.  
}

\FullConference{The 34th International Cosmic Ray Conference,\\
		30 July- 6 August, 2015\\
		The Hague, The Netherlands}

\begin{document}

\section{Introduction}

The optical system of the Large Size Telescope (LST) is an active optical system that is made up of an extensive reflective surface and
an Active Mirror Control (AMC) system. The optics is composed of one parabolic primary
mirror and one flat focal surface. The focal length ($f$) of the parabolic dish is 28\,m
and its diameter ($D$) is 23\,m, thus the $f/D$ ratio is $\sim1.2$. The parabolic shape is selected because it is
isochronous and the arrival time information of Cherenkov light signals is used to suppress the background. The
reflective surface is made of segmented mirror facets and each facet is actively aligned with actuators,
which constitute the AMC system.

\section{Mirror Facets}

\begin{figure}[htbp]
\centering
\includegraphics[height=6.5cm]{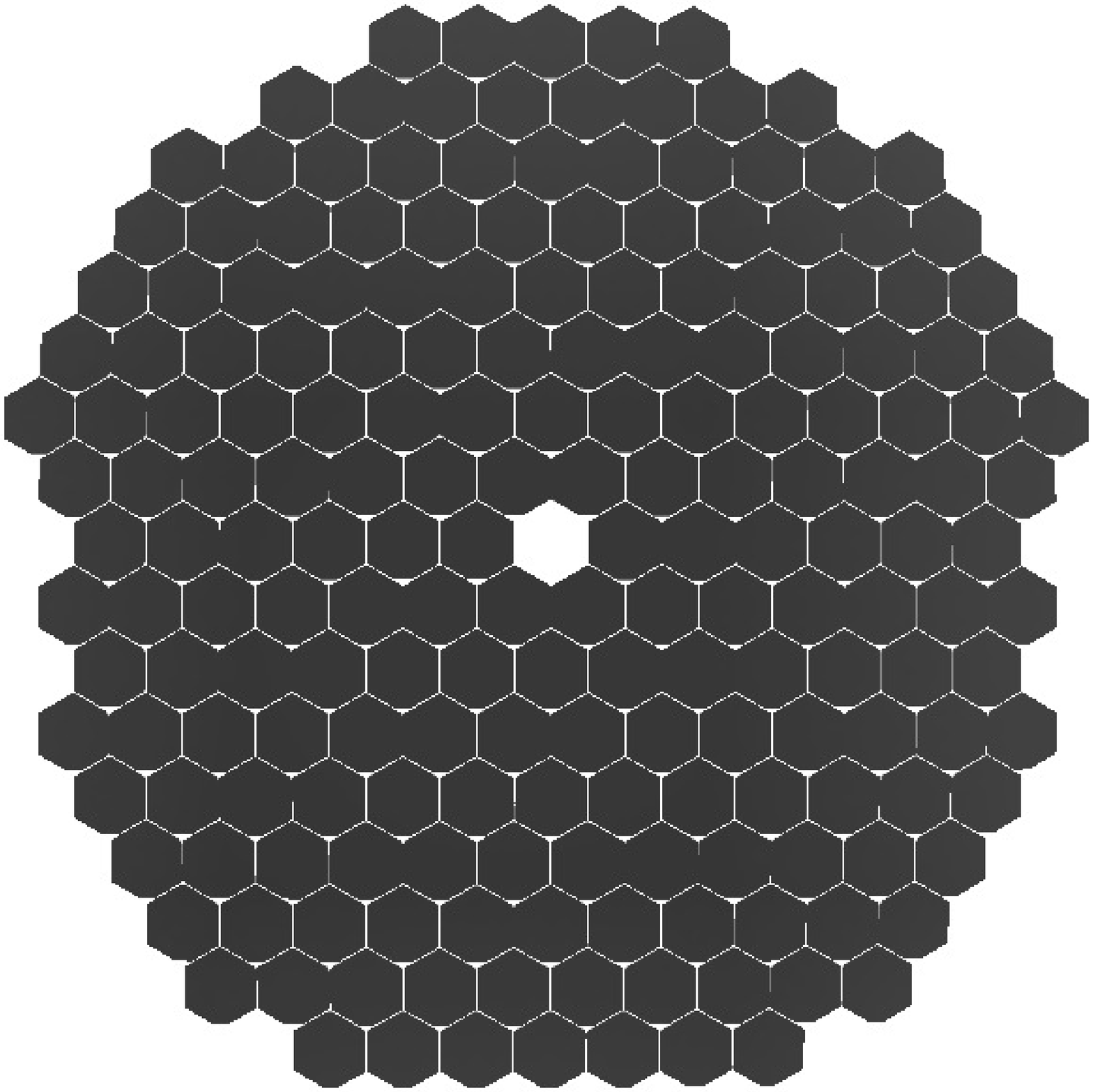}
\includegraphics[height=6.2cm]{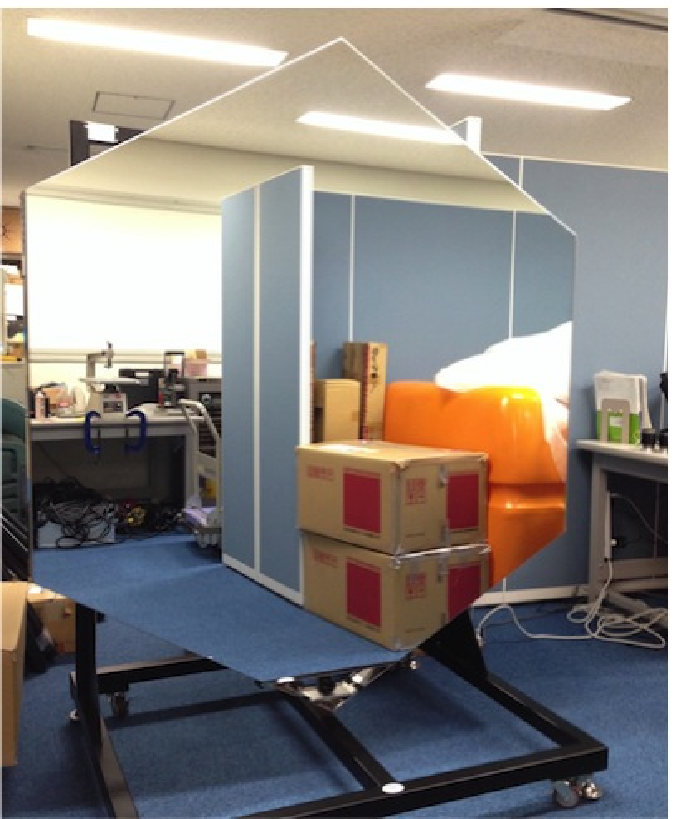}
\caption{{\bf [left]}; The arrangement of 198 mirrors on the dish space-frame structure. The hole at the centre is necessary for the installation of the calibration devices, such as the Optical Axis Reference Laser, the Star-guider camera and the camera for the PSF monitoring. {\bf [right]}: the mirror facet for LST made in Japan by a company $Sanko$. At the bottom corner, a hole is cut out from the edge of the mirror to allow for installation of a CMOS camera.}
\label{fig_mirconf}
\end{figure}

The 23\,m diameter reflector of LST is a segmented mirror telescope consisting of hexagonal facets that measure
1510\,mm flat to flat. The mirror facets are mounted at the nodes of the telescope dish. They are
arranged according to the measured characterization of each mirror (i.e.\ focal length and PSF) following
the parabolic profile of the dish. Each facet has a spherical profile with a specific focal length
that depends on its position in the reflective surface.  
The facet
at the centre of the reflector will not be installed because that part falls under the shadow of the camera.
Instead, this space is left free for the calibration devices. 198 mirror facets can be installed with the
configuration shown in Figure~\ref{fig_mirconf}-(left). The total effective reflective area, taking into account the shadows
of the mechanical structures (e.g., the photo-sensors camera, mast, etc.), is about 368\,m$^{2}$. 

The mirrors are manufactured using the cold slump technique~\cite{Par08}.
The mirror facets are fabricated by the Japanese company $Sanko$ with a sandwich structure consisting
of a soda-lime glass sheet, aluminum honeycomb and another glass sheet. 
The mirror's core is made of aluminum honeycomb because it provides an incredibly high strength-to-weight ratio.
Several small slits are made in the internal cells of the aluminum honeycomb panel, which allow 
air and any water which enters
to freely move inside. Two drainage holes are included at the bottom of the mirror,
so the water accumulated in its internal structure can be expelled. This solution prevents stationary
water in the honeycomb cells freezing during winter and destroying the mirror structure. One corner has
a cut-out (hole) of about 10\,cm from the edge to install a CMOS camera for the AMC system (see Figure~\ref{fig_mirconf}-right). 
The reflective layer of the glass mirror is coated on the surface with a protective multi-coat layer of chromium
(Cr), aluminum (Al), silicon dioxide (SiO$_2$), hafnium oxide (HfO$_2$) and again SiO$_2$. 
All five layers are produced using a sputtering deposition technology inside a vacuum chamber.
The sputtering method has the advantage of making the coating stronger (providing longer lifetime than that of normal mirrors with evaporation coating).

The mirror facets are evaluated for both the optical and the mechanical points of view. 
The radius of curvature and
PSF of each mirror facet were measured at ICRR Kashiwa in Japan, using a Phase Measuring Deflectometry (PMD) method~\cite{Kna04}.
The optical performance of the first prototypes of 29 mirror facets were measured using the PMD device. 
The PSF is defined by ``D80'', which is a diameter containing 80\% of the total reflected light.
The curvature radii of all 29 prototype mirror
facets meet the requirement (56--58.4\,m)\footnote{The focal length of the mirror facet is a half of the curvature of radius.}, while 5 mirror facets out of 29 have a PSF larger than
the requirement ($<33.3$\,mm)\footnote{This value is the PSF when using a point light source from a distance of a curvature of radius of the mirror (at $2f$). The PSF for parallel light corresponds to a half of this value.}, and have therefore been rejected. 
The manufacturing company ($Sanko$) has already identified the reasons for these failures, 
and the manufacturing process has been improved.

\begin{figure}[htbp]
\centering
\includegraphics[height=3.8cm]{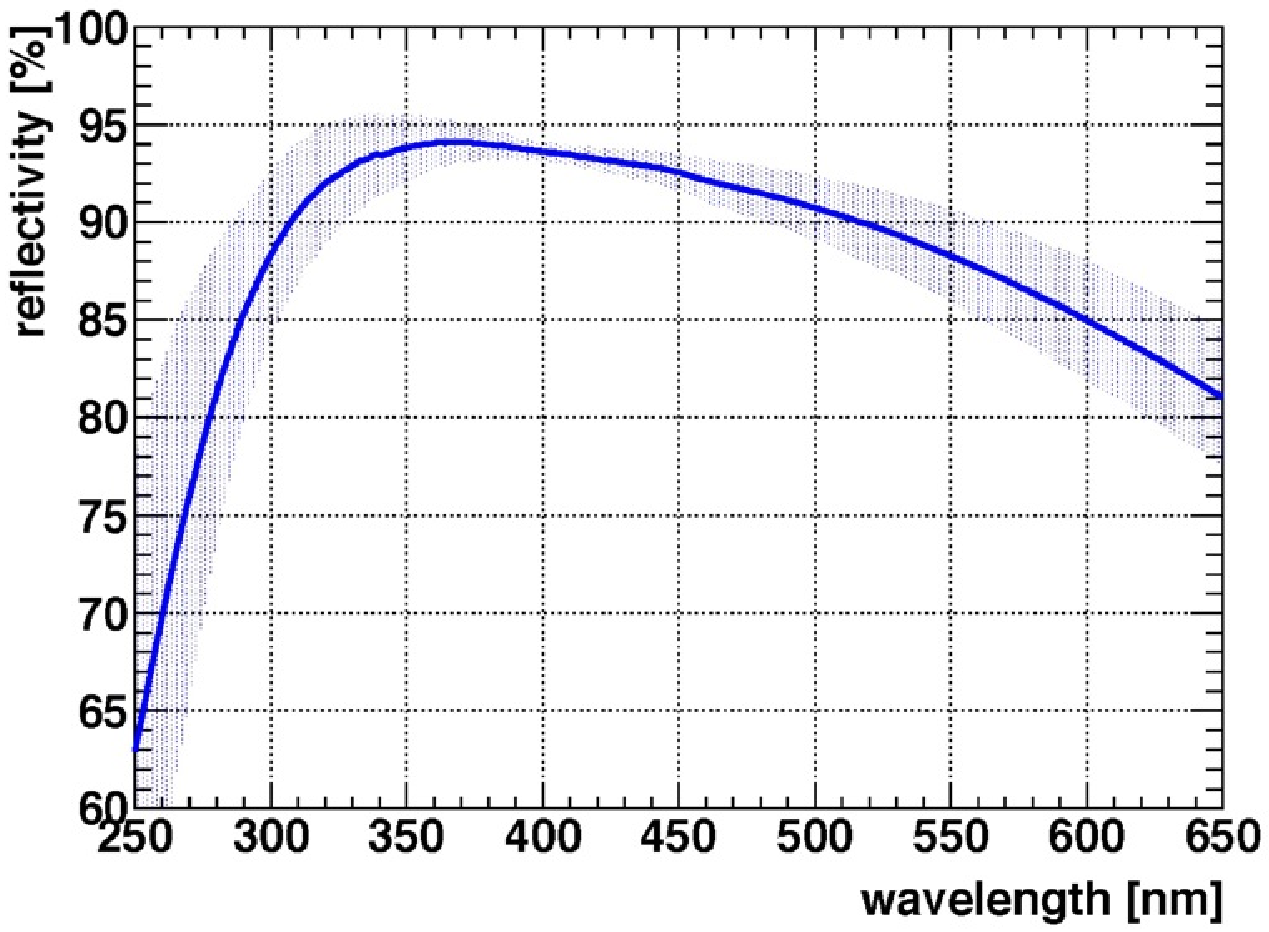} 
\includegraphics[height=3.8cm]{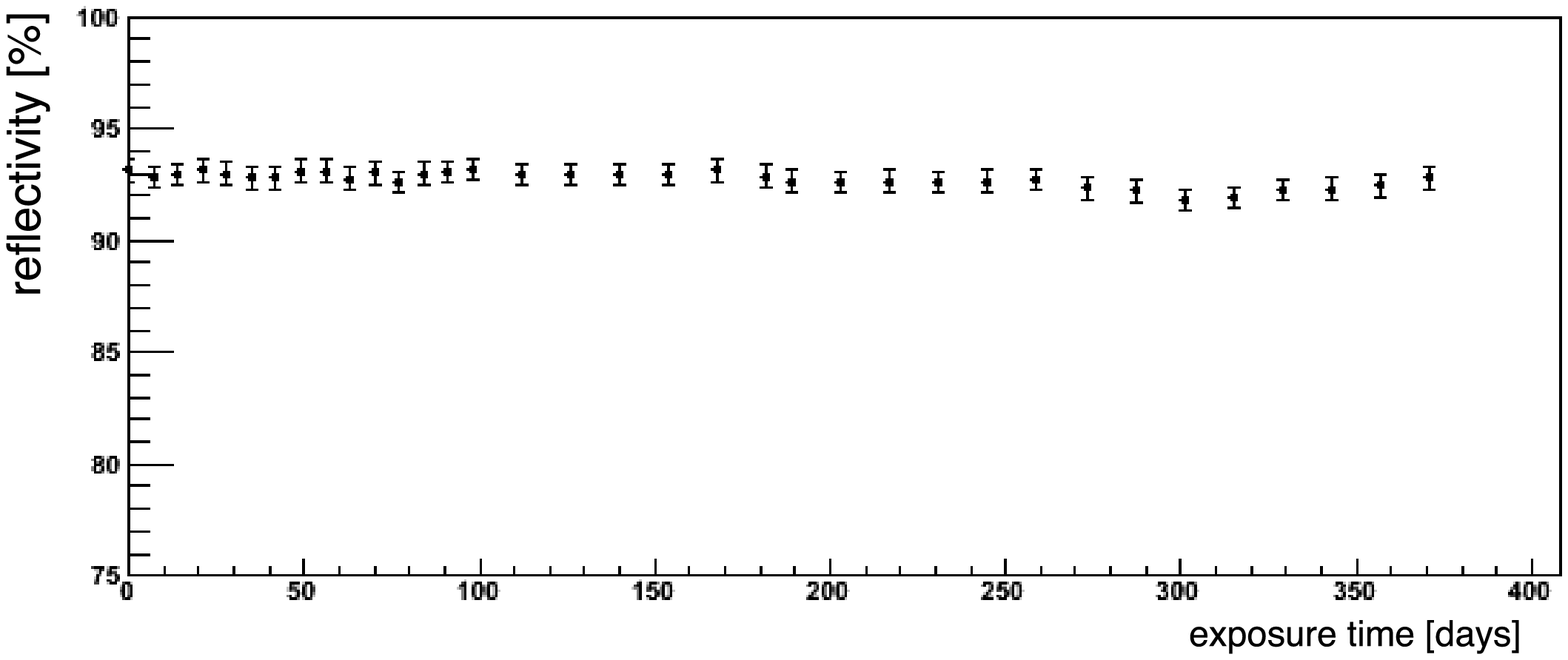}
\caption{{\bf [left]:} The spectrum of reflectance of LST mirrors with the five-layer coating averaged over the 29 samples. The error band represents 1-$\sigma$ standard deviation among the data-set. It reaches 94\% reflectance at the peak, which matches the Cherenkov light spectrum on the ground. {\bf [right]}: History of the reflectivity at 370\,nm for a sample  with the five-layer coating exposed outside in Ibraki, Japan. The reflectivity has been measured every one or two weeks.} 
\label{fig_ref}
\end{figure}

The reflectivity of the 29 prototype mirrors was measured in the wavelength range from
250\,nm to 650\,nm. Figure~\ref{fig_ref} shows the reflectivity spectrum in this range, averaged over the 29
samples. The spectrum reaches 94\% at its peak of 370\,nm and keeps more than 90\% reflectivity between
310\,nm and 510\,nm. Thus, the simple average reflectivity from 300\,nm to 550\,nm corresponds to 92.1\%,
well above the requirement. Light scattered from the surfaces of small mirror samples with the
same five-layer sputtering coating has been measured using the CTA Mirror Test Facility in Olomouc, Czech Republic.
These measurements have shown that the percentage of scattered light is very low, at less than 1\%.
Even after some durability tests,
such as abrasion and climate chamber tests, the scattering component did not increase by more than
1\%.
Several small mirror samples with the five-layer coating have been exposed at CTA candidate sites as well as some locations in Japan to check the durability of the mirror coating. 
The sample which has been exposed in Ibaraki, Japan, is being regularly monitored every one or two weeks.
It has been exposed for about one year (as of now, in 2015 June), the reflectivity was only degraded by about 1\% even in the UV range (at 370\,nm shown in Figure~\ref{fig_ref}-right).

\begin{figure}[htbp]
\centering
\includegraphics[height=4.3cm]{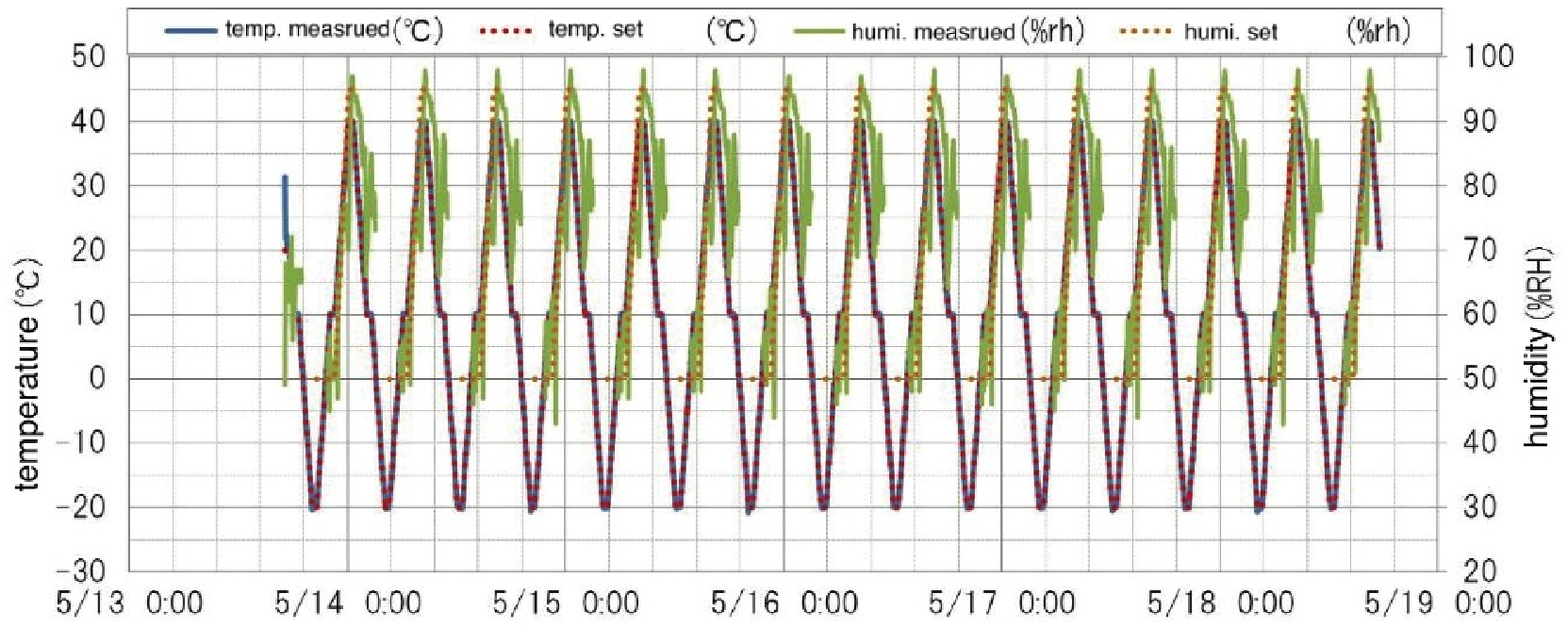} 
\includegraphics[height=4.3cm]{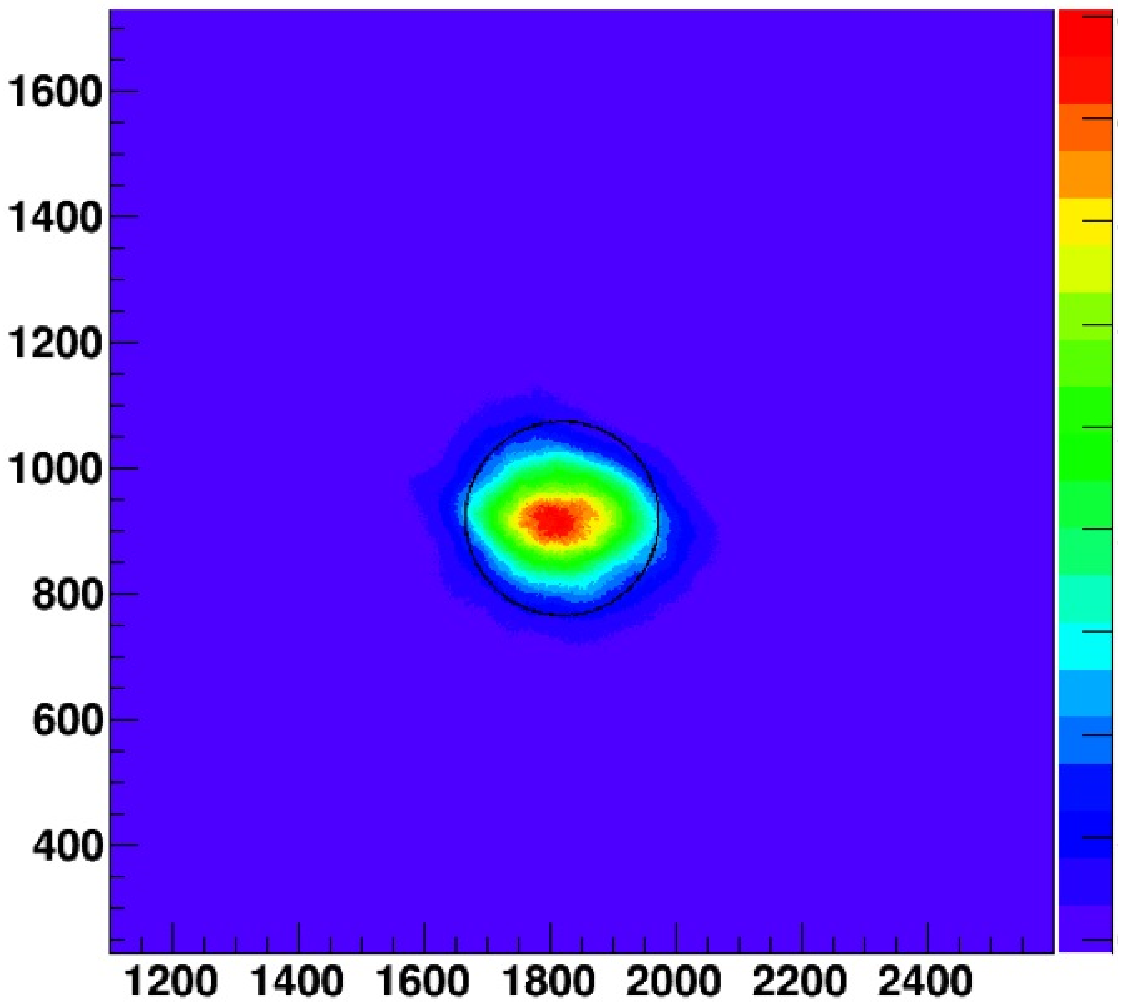}
\caption{{\bf [left]:} History of temperature and humidity during 5 days of the temperature cycle tests.
The humidity was also controlled when the temperature is above $10^{\circ}$C so that humidity can remain high.
{\bf [right]}: a spot image in linear scale at 2f position as a reflective image of a point light source after a temperature cycle test with 15 cycles from $40^{\circ}$C to $-20^{\circ}$C. The PSF, D80, is indicated by a black circle, corresponding 29.1\,mm in diameter. Numbers in both x-axis and y-axis represent pixel numbers of the imaging camera. 311.5 pixels corresponds to 30\,mm for each axis. }
\label{fig_temp}
\end{figure}

The mechanical qualities of the mirror facets are studied with extensive environmental cycling, which 
entails considerable variations of the temperature, humidity and 
external conditions such as might be expected at the CTA sites.
Temperature
cycle tests were performed for 25 days in total using a climate chamber in Matsudo, Japan, for several mechanical mirror samples\footnote{The mechanical samples have the exact same mechanical structures with the final design of the mirror, but no reflective layer was coated on the surface.} and a few real mirrors.
The tests were performed in a dry (no humidity control) condition for 10 days and 
in damp conditions (with suitable humidity control) for 10 and 5 days in total.
Inspection of the mirrors after all the tests revealed no problems with the glue or mirror structure.
Figure~\ref{fig_temp}-(left) shows records of temperature and humidity during the last 5 days of the tests.
During the test, the temperature was changed from $40^{\circ}$C to $-20^{\circ}$C with 15 cycles over 5 days.
Figure~\ref{fig_temp}-(right) shows the image of a point source of light obtained at the 2f position after the mirror had undergone temperature cycling.
The result showed no degradation in the PSF, which has a D80 of $\sim$29\,mm at 56.04\,m, consistent with the value measured before the temperature cycle test.

\section{Active Mirror Control System}

The purpose of the AMC system is to correct for the deformations of the dish and camera support structure 
that affect the alignment of the mirrors with the camera and change the focal distance.

\subsection{Actuator}
\begin{figure}[htbp]
 \begin{minipage}{0.53\hsize}
  \begin{center}
   \includegraphics[height=45mm]{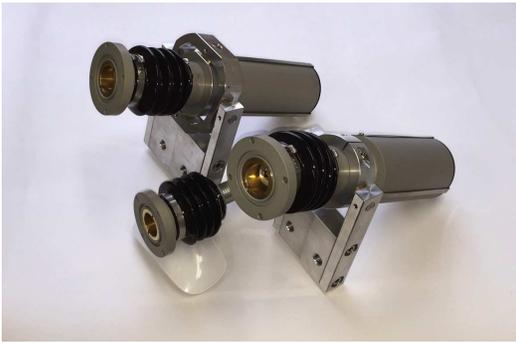}
  \end{center}
      \vspace{-5mm}
  \caption{Two mounting points including the actuators with a single and a double gimbal, and a fixed support}
   \label{fig_act}
 \end{minipage}
  \begin{minipage}{0.44\hsize}
  \begin{center}
   \includegraphics[height=45mm]{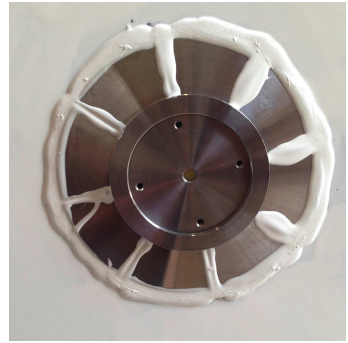}
  \end{center}
    \vspace{-5mm}
  \caption{the pads glued on the back side of the mirror for the fixation to the mounting points}
     \label{fig_pad}
 \end{minipage}
\end{figure}

The two actuators and the universal joint of each mirror facet are respectively fixed to three neighboring main joints
of the dish space frame, via special plates, called ``Interface Plates'' (see details in the next section). 
Two mounting points include actuators, while the third has a fixed support. In order to minimize the mechanical 
stress on the mirror, one actuator has only one degree of freedom (single gimbal), whereas the
other one has two degrees of freedom (double gimbal) (Figure~\ref{fig_act}). All three points have 
an universal joint where they are fixed to the mirror, which are connected to stainless steel pads glued on the rear 
of the mirror facet (Figure~\ref{fig_pad}).

The actuators for the LST have been designed by the University of
Zurich based on the MAGIC AMC by the ETH Zurich. The developments and
validations of the actuators have been performed by the University of Zurich and the University of Tokyo.
The actuators for LST are required to have a maximum moving force and
survival force respectively of greater than 700\,N and 9000\,N. 
The actuator consists of the linear movement attached to a spindle which moves a nut connected to a piston, 
thus transforming rotation into translation. The piston position is read out by a set of Hall sensors and
a magnetic angular encoder that ensure absolute safe positioning in case of power-cut. 
The actuators are self-locking, so idle power consumption is
dramatically reduced and an UPS is not needed for either the actuator or the controller. The actuators
use industry standard high efficiency 24VDC power supplies. 
An IP67 rated plug on the
actuator side allows easy and fast connecting and disconnecting. All the cables are UV, ozone and
ammonia resistant, and they are water-proofed and suited for a temperature range of $-40^{\circ}$C to $+90^{\circ}$C.
The actuators support commands such as moving
to an absolute position or by a given number of steps. 
The commands are sent from the control PC
using the wireless communication standard IEEE 802.15.4, which is adopted in applications that require a low
data rate, long battery life and secure networking. 
Each actuator has an unique 64-bit
MAC address, linked to a serial number printed on its casing. 
Any individual actuator is addressed by its unique address or by broadcast commands. The
broadcast mode allows commands to be sent simultaneously to all the actuators in a given network ID
(e.g. move all the mirrors to a specific position, previously stored in the internal look-up table, according
to the elevation angle of the dish). The upload of the look-up tables and even the firmware upgrades
can also be realized via wireless, so no physical access to the hardware is required in most occasions.

\subsection{CMOS camera}
\begin{figure}[htbp]
\centering
\includegraphics[height=4.cm]{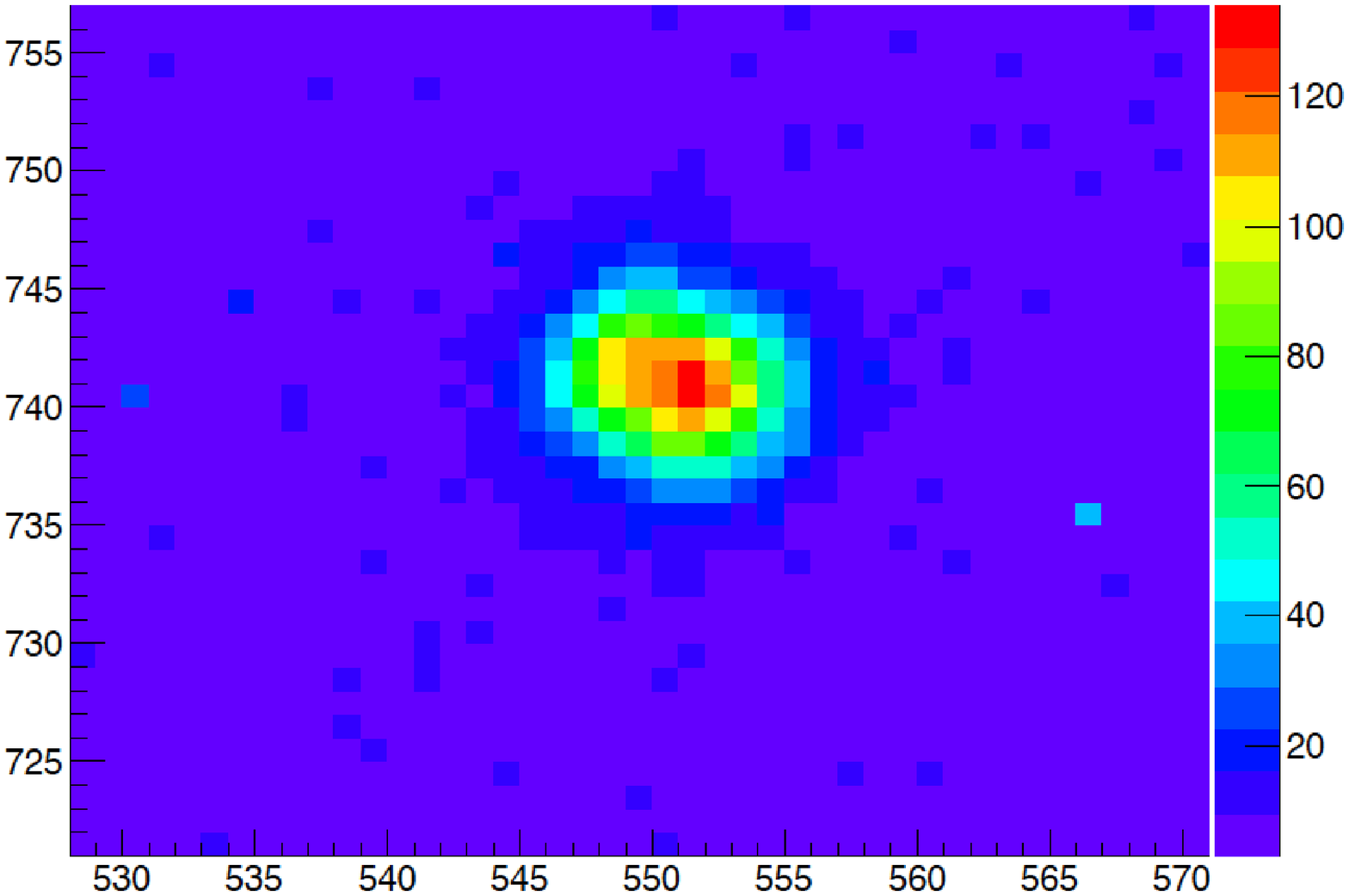}
\includegraphics[height=4.cm]{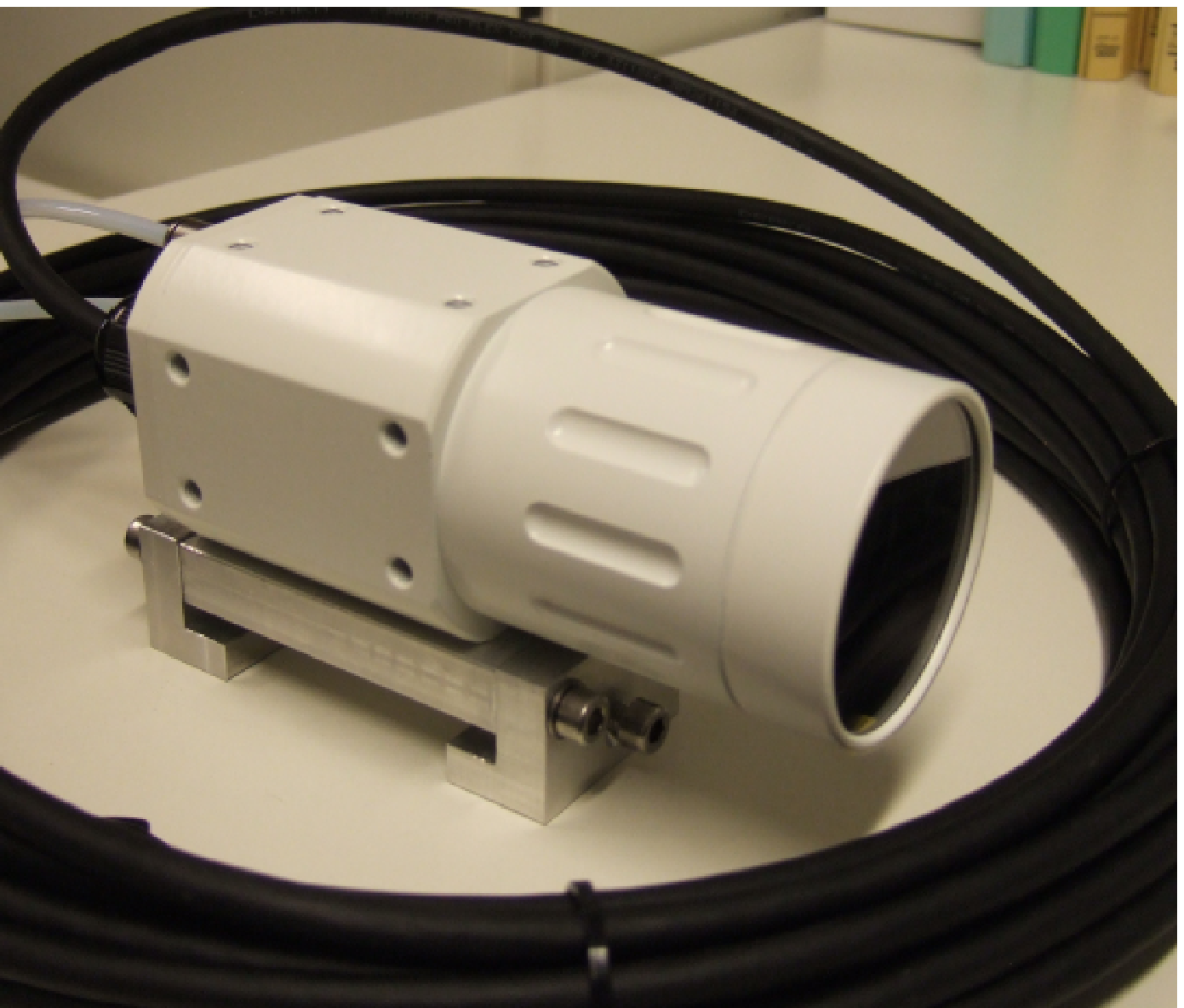}
\includegraphics[height=4.cm]{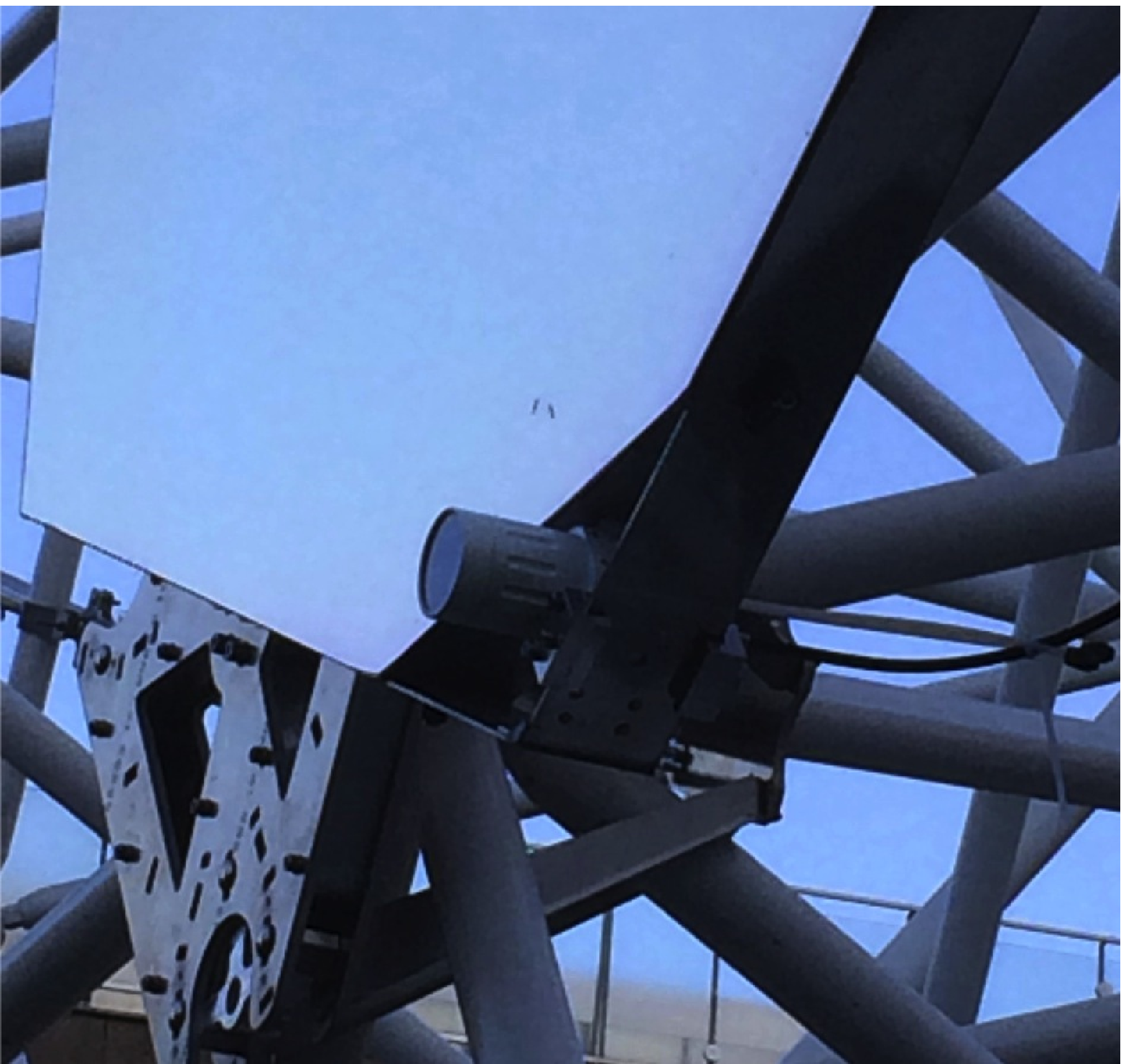}
\caption{{\bf [left]}: A spot image of the near infrared laser taken with the CMOS camera at a 28\,m distance. Numbers in both x-axis and y-axis represent pixel numbers of the CMOS camera. {\bf [middle]}: The CMOS camera packed with IP68 case. {\bf [right]}: A CMOS camera package for AMC mounted in the cut-out corner of a mirror facet.}
\label{fig_cmos}
\end{figure}

The reference of each mirror direction is defined by the spot of the Optical Axis Reference Laser (OARL), whose wavelength is in the near infrared region. 
Each mirror facet carries a CMOS camera, which measures the position of the light spot of the OARL on the target of the telescope camera in order to identify the current mirror direction with respect
to the optical axis. 
A commercial product, Imaging Source (DMK 42AUC03), has been selected for the CMOS camera.
The CMOS camera has $1280\times960$ imaging pixels with a 8-bit grayscale dynamic range.
The camera is equipped with a lens of 25 mm focal length, leading to a $10.75^{\circ} \times 8.06^{\circ}$ for the field of view of the camera.
Therefore, one pixel size is about 30\,arc-second, corresponding to 4.2\,mm on the LST focal plan at 28\,m.
Figure~\ref{fig_cmos}-(left) shows an example of the spot image of the near infrared laser taken with the CMOS camera at a 28\,m distance in the dark condition inside a building.
The laser has its beam divergence of $0.6\times0.3$\,mrad (at 1\,m), 3\,mW, and 785\,nm. The laser spot on the screen was illuminated from a 28\,m away.
Based on 100 spot images, the accuracy of the spot positional determination was estimated at 0.052 pixels, corresponding to 1.6\,arc-second.
Each CMOS camera and lens are packed into an IP68 case with a transparent face (Figure~\ref{fig_cmos}-middle) and mounted in the cut-out corner of a mirror as shown in Figure~\ref{fig_cmos}-(right).
The camera is connected with a board computer via a Gigabit Ethernet cable.

\subsection{AMC mirror alignment}
The correct position of the OARL in an image of each CMOS camera is measured during the initial
alignment and stored in a database. During the telescope operations, the AMC alignments will be performed
in two steps. When the telescopes move to a new observation target, the mirrors will be adjusted
based on look-up tables, which store the correct position of each mirror according to the elevation angle.
However, the look-up tables are pre-defined and do not consider all the deformations of the telescope
structure. More precise mirror alignments with fine tuning will be performed in a basis of the OARL
position captured by the CMOS camera on each mirror facet. The AMC computers analyze the images
and calculate a right position for each actuator. The method allows to adjust several facets together and
complete the adjustments in a short time without disturbing the data-taking. The alignment is periodically
performed in cycles of a few minutes while tracking the source.

\section{Interface plate}

\begin{figure}[htbp]
\centering
\includegraphics[width=5cm]{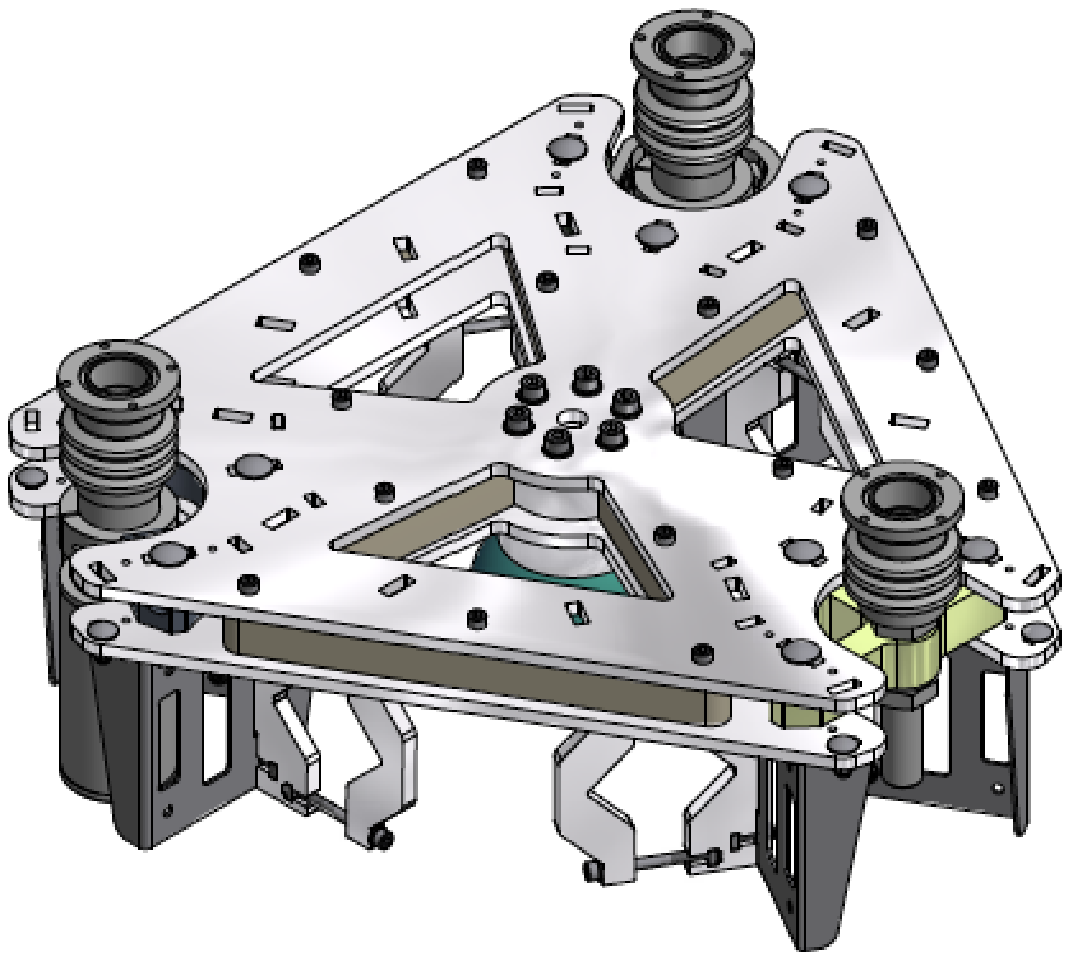}
\includegraphics[width=5cm]{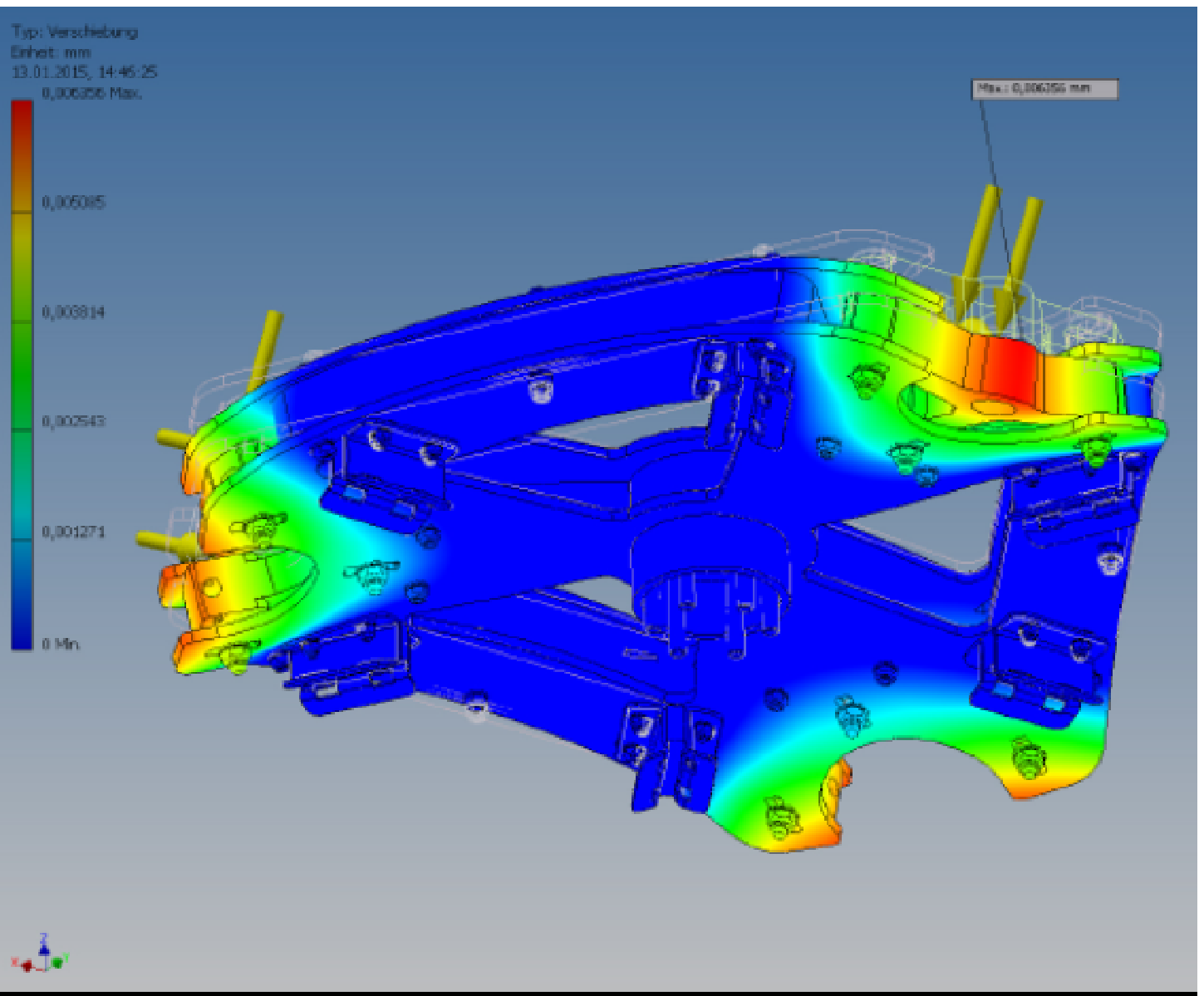}
\caption{{\bf [left]}: The drawing of the interface plate. {\bf [right]}: Two-dimensional FEA simulations of the interface plate displacement with a symmetric external load of 750N along z axis. }
\label{fig_IFplate}
\end{figure}

An ``Interface Plate'' connects three mounting joints, each from a different mirror, 
to a single main joint.
This interface is very
complex because it must be stiff also under dynamic loads like wind gusts, and at the same time, it
has an adjustable mechanism for connecting the three mounting joints, whose relative position changes
depending on the mirror's location on the dish. 
Figure~\ref{fig_IFplate} shows the design of the plate.
The plate is supported by the six surrounding tubes
to distribute the stresses as much as possible. 
Hereafter, the verification of the mechanical stiffness of both options is
presented.

\begin{figure}[htbp]
\centering
\includegraphics[width=7cm]{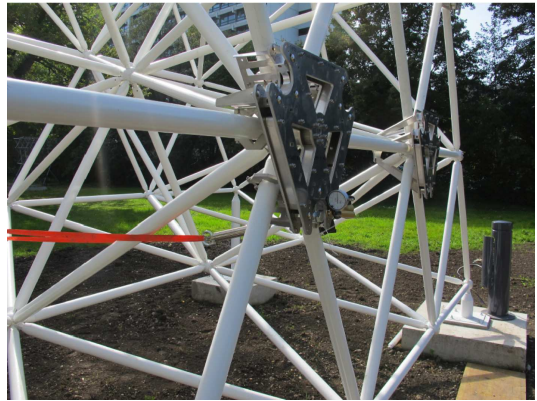}
\caption{The set-up adopted to apply an external load to the interface plate in the direction of the structure centre.
}
\label{fig_IFtest}
\end{figure}

A Finite Element Analysis (FEA) simulation has been done for the case of a symmetric load. Three loads of 250\,N
each have been applied along z axis in correspondence of the three corners where the actuators are
connected. The results are shown in Figure~\ref{fig_IFtest}. 
The maximum displacement occurs in the same regions where the loads are applied and it is $7\,\mu{\rm m}$. 
The interface plates were installed on the bottom of the test structure.\footnote{The test structure is a segment of the LST dish (roughly 1/8 of the final structure) assembled in the garden of the Max-Plank Institute for Physics in Munich.} 
An external load,
measured by a dynamometer, was applied with an adjustable belt, as shown in Figure~\ref{fig_IFtest}. 
A load of 500\,N was applied at one corner of the interface plate under test from the rear of the structure
(nearly horizontally) to check its stiffness under extreme wind condition facing the dish. Note that a load
of 500\,N is basically double what every actuator is subjected to in the presence of a wind of 70 km/h, which is
more testing than the worst condition during normal operation. Moreover, the asymmetry of the applied load is
also extreme and assures that the study is very conservative. The displacements of the interface plates
were measured with a gauge fixed to a nearby tube.
The measured displacements of the plate at the fixation points were about 30\,$\mu$m.
The displacements were caused by not only the deformation of the plate itself but also the deformation of the space frame structure.
Additional FEA simulation was performed to estimate the displacements including the space frame structure in the same conditions of the field test. The results were consistent with the measurements.

The next study will optimize how to clamp the interface plates on carbon-fiber-reinforced plastic (CFRP) tubes, as the test structure is made of steel tubes. Some small modifications in the clamping mechanism could be necessary to ensure that CFRP surface is not damaged.

\section{Summary}

The LST optics is composed of one parabolic primary mirror 23\,m in diameter and 28\,m focal length. The reflector dish is segmented in 198 hexagonal, 1.51\,m flat to flat mirrors. The surface and reflectivity were measured for the first 29 prototype mirrors, and temperature cycle tests were also performed for a few mirrors.
The AMC system will actively align mirrors during the operation by actuators.
Each mirror facet carries a CMOS camera with a lens of 25 mm focal length
which measures the position of the light spot of the optical axis reference laser on the target of the telescope camera. 
The two actuators and the universal joint of each mirror facet are respectively fixed to three neighboring main joints of the dish space frame, via interface plate. The mechanical stiffness of the plate were verified by both FEA simulations and field tests.

\section{Acknowledgement}
We gratefully acknowledge support from the agencies and organizations
listed under Funding Agencies at this website: http://www.cta-observatory.org/, 
especially the supports by JSPS, ICRR. U.\ Tokyo, Max-Planck-Society,
BMBF, U.\ Hamburg, CNPq, FAPERJ, FAPESP, and U.\ Zurich.

\end{document}